\documentclass[prl]{revtex4}
\usepackage{graphicx}
\usepackage{dcolumn}
\usepackage{bm}
\begin{document}

\author{N. Oeschler,$^1$ R.\,A. Fisher,$^1$ N.\,E. Phillips,$^1$ J.\,E. Gordon,$^2$ M.\,L. Foo,$^3$ and R.\,J. Cava$^3$}
\affiliation{$^1$Lawrence Berkeley National Laboratory and
Department of Chemistry, University of California, Berkeley, CA,
94720, USA} \affiliation{$^2$Physics Department, Amherst College,
Amherst, MA, 01002, USA} \affiliation{$^3$Department of Chemistry,
Princeton University, Princeton, NJ, 08544, USA}

\title{Heat Capacity of Na$_{0.3}$CoO$_{2}$$\cdot$1.3H$_{2}$O, a New Two-Gap Superconductor: \\ Comparison with the Heat Capacity of MgB$_2$}

\date{\today}
\begin{abstract}The superconducting-state heat capacity of
Na$_{0.3}$CoO$_{2}$$\cdot$1.3H$_{2}$O shows unusual, marked
deviations from BCS theory, at all temperatures. At low
temperatures the heat capacity has the $T^2$ dependence
characteristic of line nodes in the energy gap, rather than the
exponential temperature dependence of a fully gapped, conventional
superconductor. At temperatures of the order of one fifth of the
critical temperature and above, the deviations are strikingly
similar to those of MgB$_2$, which are known to be a consequence
of the existence of substantially different energy gaps on
different sheets of the Fermi surface. A two-gap fit to the
Na$_{0.3}$CoO$_{2}$$\cdot$1.3H${_2}$O data gives gap amplitudes of
45\% and 125\% of the BCS value, on parts of the Fermi surface
that contribute, respectively, 45\% and 55\% to the normal-state
density of states. The temperature of the onset of the transition
to the vortex state is independent of magnetic field, which shows
the presence of unusually strong fluctuations.
\end{abstract}

\maketitle
\section{Introduction}
Among the unusual superconductors discovered in recent years, two, MgB$_2$ in 2001 \cite{nag01} and
Na$_{0.3}$CoO$_{2}$$\cdot$1.3H$_{2}$O in 2003 \cite{tak03}, are superconducting at ambient pressure. Both are of special interest in connection with unusual features of the electron-pairing mechanism, and in both cases the heat capacity ($C$) shows unusual, strong deviations from BCS theory \cite{bcs} that are related to these features. At temperatures ($T$) that are well below the critical temperature ($T_c$) the deviations are different in the two cases, and suggest a difference in the electron pairing; at intermediate and high temperature the deviations are very similar, and show the presence of two distinctly different energy gaps.\\The BCS theory of 1957 showed that high values of $T_c$ were favored by small atomic mass. Given the ensuing search for superconductivity in compounds of light elements, including B, and the commercial availability of MgB$_2$ for many years, the late date of the (apparently accidental) discovery of superconductivity in MgB$_2$ was itself a surprise. However, the initial intense interest generated by the discovery derived from the fact that, although the isotope effect \cite{iso} suggested that MgB$_2$ was a "conventional" BCS superconductor with phonon-mediated electron pairing, $T_c$ = 39~K seemed too high for that mechanism. The BCS phonon mechanism has been confirmed by theoretical work \cite{an01,kor01,liu01,cho02a,cho02b,gol}, which, in concert with a variety of experimental measurements including heat-capacity measurements, has also shown that MgB$_2$ is an example of multi-gap superconductivity. The possibility of multi-band, multi-gap superconductivity had been recognized theoretically \cite{smw} in 1959 and a number of the properties predicted \cite{smw,kres}, but MgB$_2$ is the first example to show the properties so clearly and in such detail. For the superconducting-state, conduction-electron contribution ($C_{es}$) to $C$ the relevant theoretical predictions for a two-gap superconductor are: 1) for any realistic interband coupling the two gaps will open at a common $T_c$; 2) one gap must be smaller than the BCS gap and the other greater; 3) at low temperature $C_{es}$ is determined by the small gap; 4) the discontinuity in $C_{es}$ at $T_c$ is smaller than the BCS value. For MgB$_2$, $C_{es}$ conforms to all of these predictions and is accurately represented by a two-gap model that is in quantitative agreement with both spectroscopic measurements \cite{spec}
and theoretical calculations \cite{cho02a,cho02b,gol}. MgB$_2$ is well established as a two-gap superconductor, and $C_{es}$ can be taken as a model for comparison with other superconductors that might have more than one gap.\\The superconductivity of
Na$_{0.3}$CoO$_{2}$$\cdot$1.3H$_{2}$O is of particular interest for comparison
with that of the cuprates: The 1986 discovery \cite{bm} of superconductivity in the layered perovskite cuprates, for which $T_c$ reaches 133~K \cite{sch}, raised the question of whether high-$T_c$ superconductivity might be found in similar structures with transition-metal ions other than Cu. Co was recognized as an interesting candidate almost immediately, but the superconductivity of
Na$_{0.3}$CoO$_{2}$$\cdot$1.3H$_{2}$O, with a relatively low $T_c$, about 5~K, was not discovered until 2003 \cite{tak03}. In the cuprates the Cu ions are in corner-sharing O octahedra or pyramids, in an approximately square array. In the parent undoped Mott insulator the Cu ions are ordered antiferromagnetically, and antiferromagnetic spin fluctuations are thought to have a role in the electron pairing in the hole-doped superconducting phases. In Na$_{0.3}$CoO$_{2}$$\cdot$1.3H$_{2}$O the Co ions are at the centers of
$\it{edge}$-sharing O octahedra, in a $\it{triangular}$ array, which produces
frustration of the antiferromagnetic interaction that is expected to affect the superconductivity.\\
Theoretical work suggests that the electron pairing in Na$_{0.3}$CoO$_{2}$$\cdot$1.3H$_{2}$O is "unconventional", and perhaps unique. Experimental work has been limited by the complicated synthetic
procedures, the sensitivity of the superconductivity to
stoichiometry, and the relative instability of the superconducting
material. Only a few measurements that give clues to the nature of the pairing state have been made, and some of the results are contradictory: NQR and NMR spin-lattice relaxation rates \cite{nqmr} and $\mu$SR measurements \cite{kan} suggest line nodes in the energy gap, but a coherence peak in NQR measurements suggests an isotropic gap
\cite{gap}. $C_{es}$ gives information about the symmetry of the gap: For conventional superconductors an exponential decrease of $C_{es}$ corresponds to a fully gapped Fermi surface. For a superconductor with nodes in the
gap a power-law temperature dependence is expected, with the
exponent determined by the form of the nodes. Point nodes give $C_{es}\propto T^3$; line nodes give $C_{es}\propto T^2$. A number of heat-capacity measurements have been reported \cite{sph1,sph2}, but in many cases the interpretation of the data has been limited by substantial contributions from magnetic impurities or a lack of low-temperature data. New measurements of the heat capacity of Na$_{0.3}$CoO$_{2}$$\cdot$1.3H$_{2}$O, from 0.8 to 33~K and in magnetic fields ($B$) to 9~T,
are reported here and compared with previously published results for MgB$_2$.
\section{Heat Capacity of MgB$_2$}
Several measurements of the heat capacity of MgB$_2$ show the presence of excitations at energies below the gap energy of a BCS superconductor, i.e., a second small gap on a part of the Fermi surface \cite{cmgb,phyc}. Figure~1 shows comparisons of $C_{es}$ data \cite{phyc} with a two-gap model \cite{2gap} and with the BCS theory for the weak-coupling limit, as $C_{es}/\gamma T$ vs
$T/T_c$, where $\gamma$ is the coefficient of the conduction-electron, normal-state contribution ($C_{en}$ = $\gamma$$T$) to $C$. (For this sample $T_c = 38.9$~K and
$\gamma$ = 2.53~mJ~mol$^{-1}$~K$^{-2}$.) Near $T_c$, where the BCS
curve has negative curvature, the data show positive curvature,
which is characteristic of strong coupling. However, for
strong-coupling superconductors the discontinuity in $C_e$ at
$T_c$ is generally greater than the BCS value; in this case it is
less. This apparent discrepancy is resolved when there are two
gaps on different sheets of the Fermi
surface that make additive contributions to $C_{es}$: The large gap determines the $\it{shape}$ of $C_{es}$ near $T_c$ but the $\it{magnitude}$ of that contribution is reduced by the fractional contribution of that sheet to $\gamma$, $\gamma_1/\gamma$, and the small-gap sheet makes only a small contribution to $C_{es}$ near $T_c$. Below $T_c/2$, $C_{es}/\gamma T$ is greater than the BCS
values, by orders of magnitude at the lowest temperatures. This
excess contribution shows the presence of excitations at energies
below that of the BCS gap, i.e., excitations across the small gap. These
features in $C_{es}$ are well described by a two-gap fit \cite{2gap},
as shown in Fig.~1. That fit is based on the $\alpha$ model \cite{alph}, a
semi-empirical extension of the BCS thermodynamics to
strong-coupling superconductors. In the weak-coupling limit of
the BCS theory the 0-K gap parameter, $\Delta(0)$, is given by
$\Delta(0)/k_BT_c = 1.764$. For a strong-coupling superconductor
$\Delta(0)/k_BT_c > 1.764$, and in the $\alpha$ model that ratio is
taken as an adjustable parameter, $\alpha\equiv\Delta(0)/k_BT_c$. For a strong-coupling superconductor the temperature dependence of the gap parameter, $\Delta(T)$, can be different from that
of the weak-coupling limit, but in the $\alpha$ model it is approximated by the same temperature dependence. The model gives the thermodynamic
properties of superconductors with
thermodynamic consistency. For strong-coupling superconductors
(e.g., Pb, for which the experimental value of $\alpha$ is 2.4) it
gives the thermodynamic properties in agreement with experiment, suggesting that the
approximation used in the temperature dependence of $\Delta(T)$ is
not a serious shortcoming. For a single-gap superconductor a
value of $\alpha$ less than that of the weak-coupling limit would
be physically unrealistic, but for one of the gaps of a two-gap
superconductor it is thermodynamically required. In Fig.~1,
$C_{es}$ is represented as the sum of contributions from two
sheets of the Fermi surface with different values of $\alpha$,
weighted by their fractional contributions to $\gamma$,
$\gamma_1/\gamma$ and $\gamma_2/\gamma$. The curve for
each contribution makes its role in determining the overall
temperature dependence of $C_{es}$ clear. This interpretation of
the heat capacity of MgB$_2$, including the values of the three
adjustable parameters, $\alpha_1$, $\alpha_2$, and
$\gamma_1/\gamma_2$, is in good agreement with a number of spectroscopic
measurements \cite{spec} and theoretical
calculations \cite{cho02a,cho02b,gol}.  The two-gap
nature of MgB$_2$ is thus well established.
\section{Heat capacity of Na$_{0.3}$CoO$_{2}$$\cdot$1.3H$_2$O}
Na$_{0.3}$CoO$_{2}$$\cdot$1.3H$_{2}$O was prepared by chemical
de-intercalation of sodium from Na$_{0.75}$CoO$_{2}$ and subsequent
washing with water \cite{foo}. A $0.5$-g sample of the resulting
powder was placed in a Au-plated Cu container with excess water
and lightly pressed to promote thermal contact. The heat capacity
was measured with a heat-pulse technique. The amount of Na$_{0.3}$CoO$_{2}$$\cdot$1.3H$_{2}$O in the water/sample mixture was determined by chemical analysis. The data were
corrected for the contributions of the container (calculated) and
the excess water (from published values of the heat capacity \cite{h2o}).\\The heat capacity
is shown as $C/T$ in Fig.~2 for $B = 0$ and 9~T. The
anomaly at 4.5~K in the zero-field data marks the transition to the superconducting state. The low-temperature upturn in the 9-T data is a $T^{-2}$ hyperfine contribution. The 9-T data do not show the evidence of a transition to the vortex state that is apparent in the data for all other fields (see Fig.~6). Evidently the sample is normal to the lowest temperature in 9~T, and $B_{c2}$ is approximately 9~T. After subtraction of the $T^{-2}$ term , $\gamma$ and the lattice
contribution to the heat capacity ($C_l$) were determined by fitting the 9-T data below 12~K and the zero-field data between 6 and 12~K with the
expression $C = C_{en} + C_l = \gamma T + B_3T^3 + B_5T^5 + B_7T^7$.
 The fit gave $\gamma = 16.10$ mJ mol$^{-1}$ K$^{-2}$,
near the high end of the range of published values, which vary by about 30\%
\cite{sph1,sph2}, and $B_3 = 0.126$ mJ mol$^{-1}$K$^{-4}$, at the low
end of the range of published values, which vary by a factor of
five. The wide range of reported values of $B_3$ may be due to inadequate corrections for the contribution of excess water.\\ There is also a
small $T^{-2}$ upturn in the zero-field data, which is barely
visible in Fig.~2. It could arise from a low concentration of
weakly interacting paramagnetic centers, but the absence of
Schottky anomalies in any of the in-field measurements (see particularly Figs.~2 and 3 where the zero-field and 9-T data are
essentially indistinguishable above 6~K) shows that the
concentration is too low to have a significant effect on the interpretation of the data.
Another possibility is that it is a hyperfine contribution from a
magnetic impurity in the container. Apart from the small $T^{-2}$
term, the lowest-temperature, zero-field data are linear in $T$.
An extrapolation of this temperature dependence to 0~K gives the
correct entropy at 6~K, which supports its validity. The extrapolation also gives a non-zero intercept
at 0~K, a normal-state-like residual $\gamma$, $\gamma_r$ =
6.67~mJ~mol$^{-1}$~K$^{-2}$. All heat-capacity measurements on
this material show evidence of a contribution of this type,
and we make the usual interpretation, which is that a fraction of the sample, $\gamma_r/\gamma$ = 0.41, does not become superconducting. For each field except 9 T the electron contribution to the heat capacity ($C_e$) was calculated by subtracting $\gamma_r$$T$, $C_l$, and the $T^{-2}$ term from $C$, and scaling the result to correspond to one mole of superconducting material. The results for 0 and 9~T, $C$(9~T) = $C_{en}$ and $C$(0) = $C_{es}$, are shown in Fig.~3. The entropy-conserving construction on $C_{es}$ gives a discontinuity, $\Delta C_{es}$($T_c$)$/T_c$ = 13 mJ mol$^{-1}$~K$^{-2}$ at $T_c$ = 4.52~K. At low temperatures $C_{es}$ shows the $T^2$ dependence characteristic of line nodes in the gap, not the exponential temperature dependence of a fully-gapped conventional superconductor. The same conclusion has been reported by Yang et al. \cite{sph2}.\\In Fig.~4, $C_{es}$ data for both Na$_{0.3}$CoO$_{2}$$\cdot$1.3H$_{2}$O and MgB$_2$ are compared with BCS theory. Above $T/T_c = 0.2$ the experimental results, and their deviations from BCS theory, are remarkably similar. Given the unusual nature of these deviations from BCS theory, and the well established explanation of their origin in MgB$_2$ as a manifestation of two different energy gaps, they show that Na$_{0.3}$CoO$_{2}$$\cdot$1.3H$_{2}$O is another two-gap superconductor. A two-gap fit to the Na$_{0.3}$CoO$_{2}$$\cdot$1.3H$_{2}$O data, similar to that for MgB$_2$ in Fig.~1, is shown in Fig.~5. The major difference between the two fits, which is required by the fact that the $\alpha$ model is based on BCS thermodynamics and cannot represent the effect of nodes in the gap, is that the fit for Na$_{0.3}$CoO$_{2}$$\cdot$1.3H$_{2}$O is made only for $T/T_c \geq 0.4$.
The dotted extension of the small-gap contribution below
$T/T_c = 0.4$ is just the difference
between the experimental data and the extrapolated large-gap
contribution. However, the small-gap contribution determined in this way, including the extension below $T/T_c = 0.4$, is very similar to the contribution of a part of the Fermi surface with a small gap and line nodes in Sr$_2$RuO$_4$ \cite{nomu}.\\
The vortex-state results, shown in Fig.~6, give evidence of unusual strong fluctuation effects, possibly related to an unusual pairing mechanism. In most superconductors the onset temperature of the transition to the vortex state is reduced with increasing field. For Na$_{0.3}$CoO$_{2}$$\cdot$1.3H$_{2}$O, to within the precision of the data, it is the same in all fields. Fluctuations do broaden the transition to the vortex state, but the magnitude of the effect in Na$_{0.3}$CoO$_{2}$$\cdot$1.3H$_{2}$O seems to be without precedent, even greater than in the cuprates, the first superconductors in which it was observed. The extrapolations of the vortex-state $C_e$ to 0 K define the magnetic field dependence of the coefficient of the temperature-proportional term, $\gamma$($B$), which is shown in the inset to Fig.~6, and compared with that for MgB$_2$ in Fig.~7. The sharp initial increase in $\gamma$ with increasing field for MgB$_2$ is associated with the closing of the small gap in small fields. For Na$_{0.3}$CoO$_{2}$$\cdot$1.3H$_{2}$O, $\gamma$ is almost linear in field and the deviations from linearity are on the order of what might be expected from anisotropy of $B_{c2}$. This difference in behavior is apparently another manifestation of the line nodes in the gap.
\section{Conclusion}
Na$_{0.3}$CoO$_{2}$$\cdot$1.3H$_{2}$O is another two-gap superconductor, with gaps that are approximately 45$\%$ and 125$\%$ of the BCS values. The parameters characterizing the two parts of the Fermi surface with different gaps are remarkably similar to those for MgB$_2$. In contrast with MgB$_2$, however, the low-temperature, superconducting-state heat capacity has a $T^2$ dependence that shows the presence of line nodes in the small gap, and unconventional supercoductivity. The field independence of the onset temperature of the transition to the vortex state shows the presence of unprecedented strong fluctuation effects, which may be associated with a novel pairing state.

%
%
\begin{figure}\begin{center}\includegraphics[angle=-90, width=\textwidth]{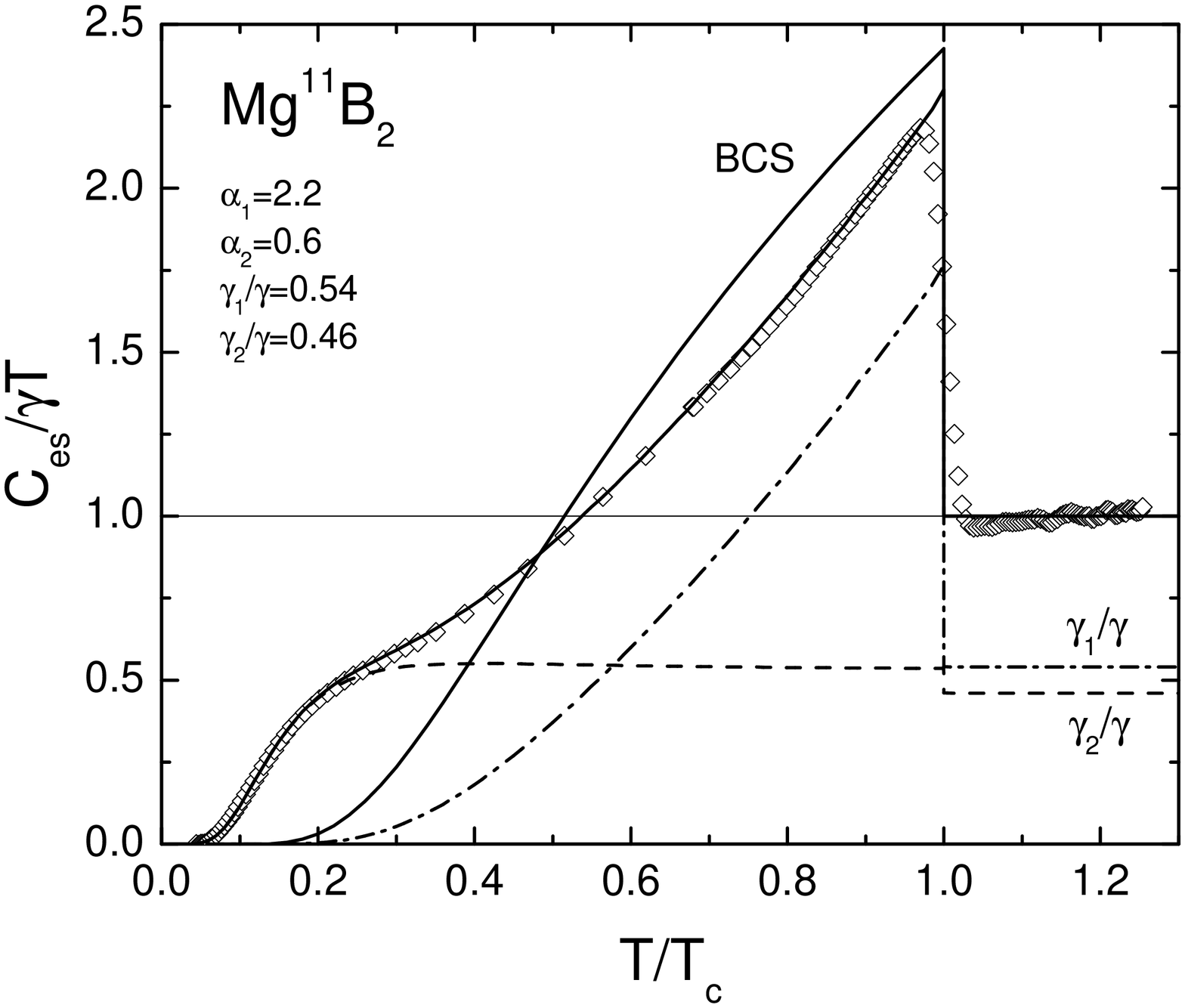}
\caption{Conduction-electron, superconducting-state heat capacity
of MgB$_2$ compared with BCS theory and a two-gap
model.}\end{center}\end{figure}
\begin{figure}\begin{center}\includegraphics[angle=-90, width=\textwidth]{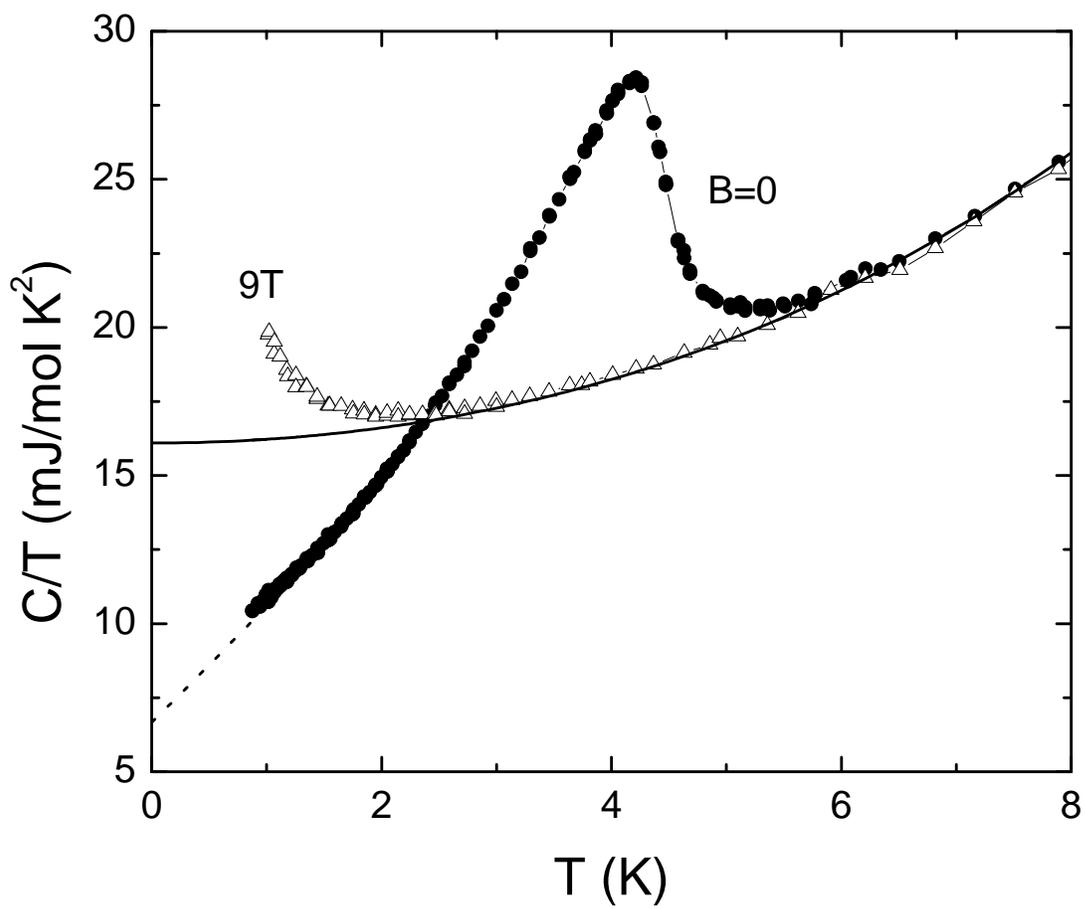}
\caption{Heat capacity of Na$_{0.3}$CoO$_2$$\cdot$1.3H$_2$O in 0
and 9 T. }\end{center}\end{figure}
\begin{figure}\begin{center}\includegraphics[angle=-90, width=\textwidth]{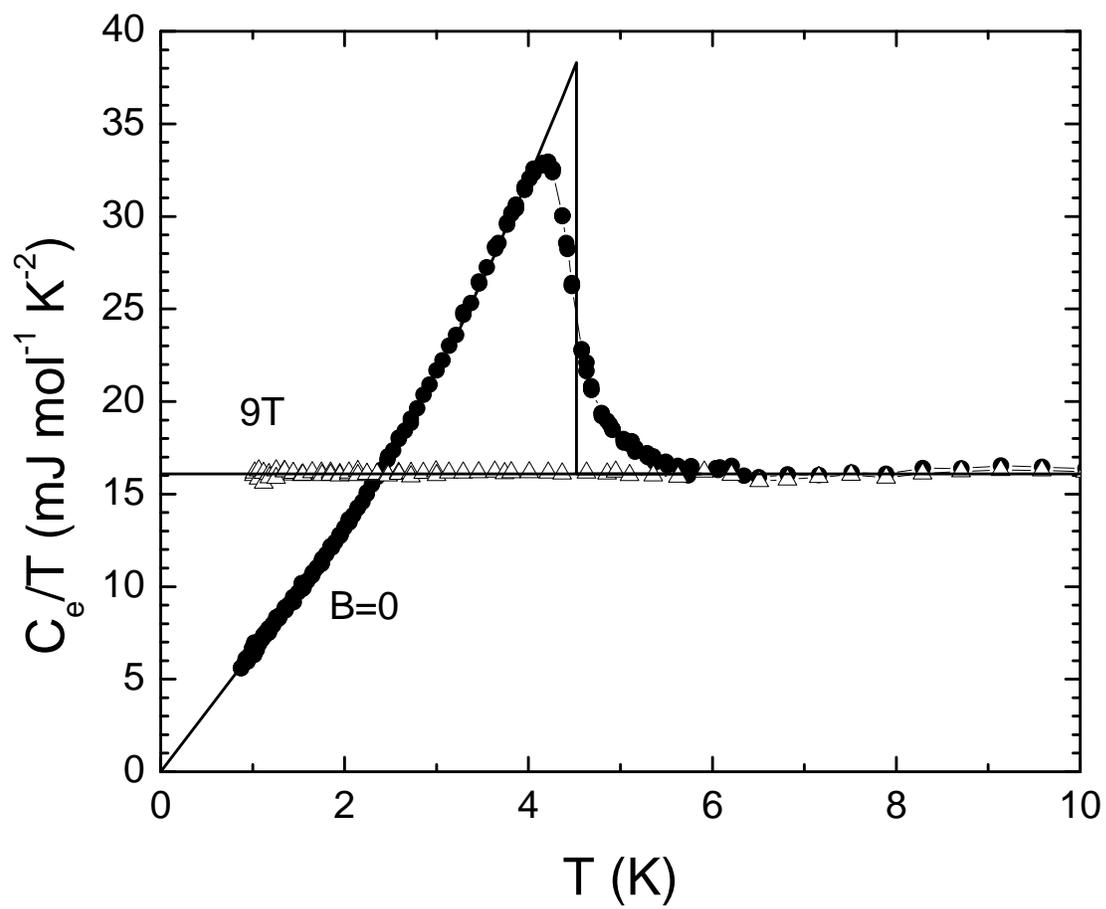}
\caption{Electron contribution to the heat capacity of
Na$_{0.3}$CoO$_2$$\cdot$1.3H$_2$O in 0 and 9
T.}\end{center}\end{figure}
\begin{figure}\begin{center}\includegraphics[angle=-90, width=\textwidth]{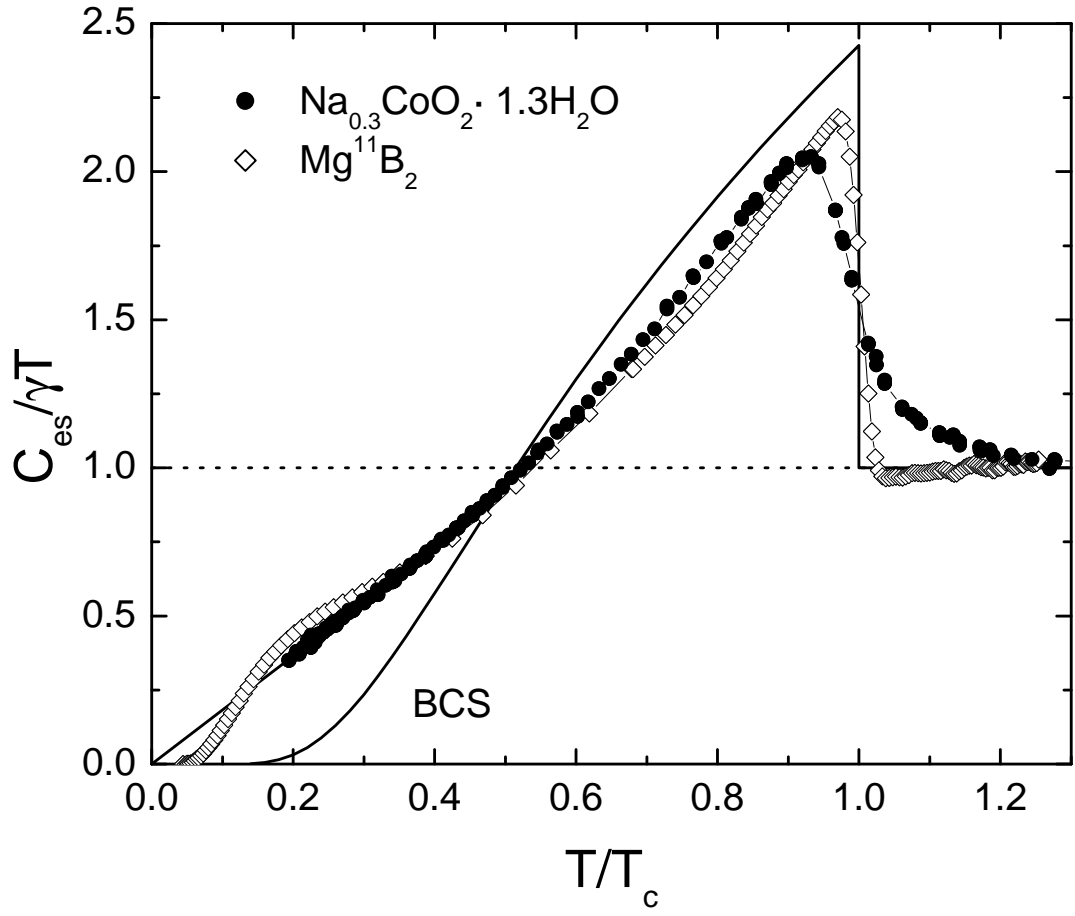}
\caption{Deviations of the heat capacities of MgB$_2$ and
Na$_{0.3}$CoO$_2$$\cdot$1.3H$_2$O from BCS
theory.}\end{center}\end{figure}
\begin{figure}\begin{center}\includegraphics[angle=-90, width=\textwidth]{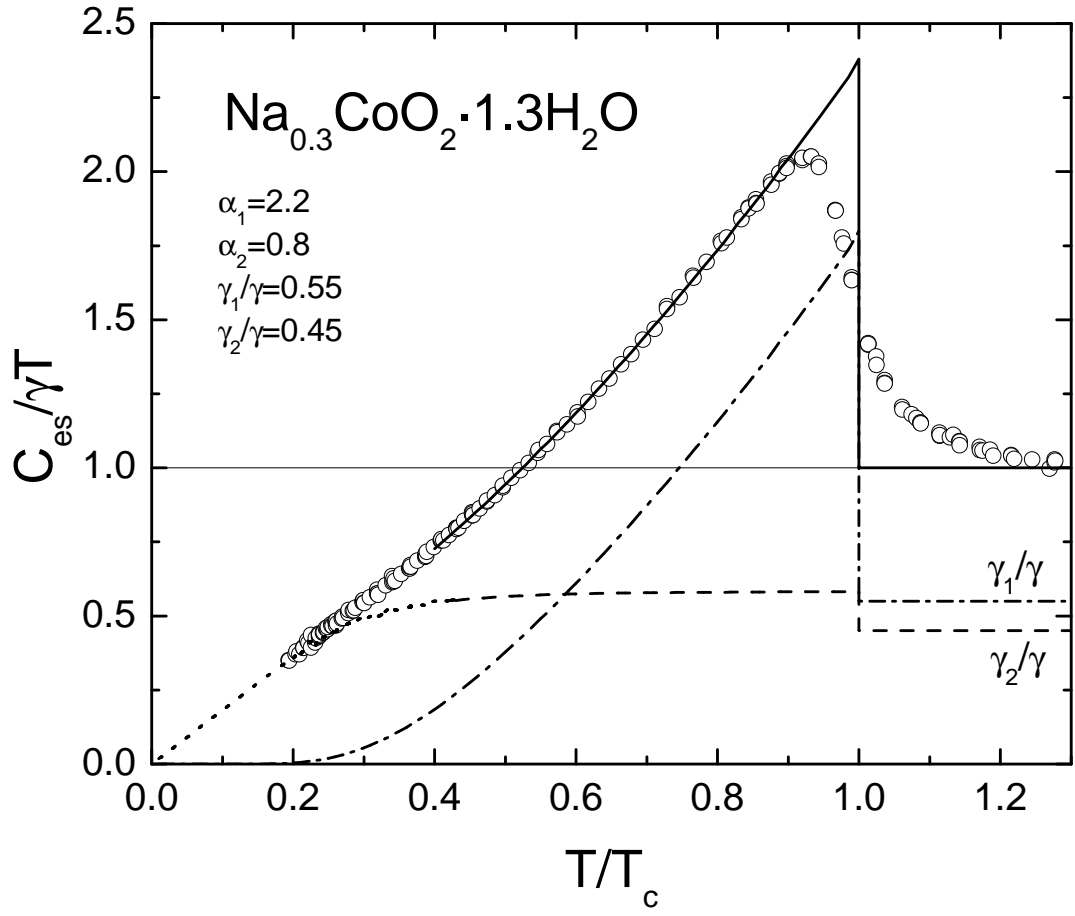}
\caption{A two-gap fit to the heat capacity of
Na$_{0.3}$CoO$_2$$\cdot$1.3H$_2$O.}\end{center}\end{figure}
\begin{figure}\begin{center}\includegraphics[width=\textwidth]{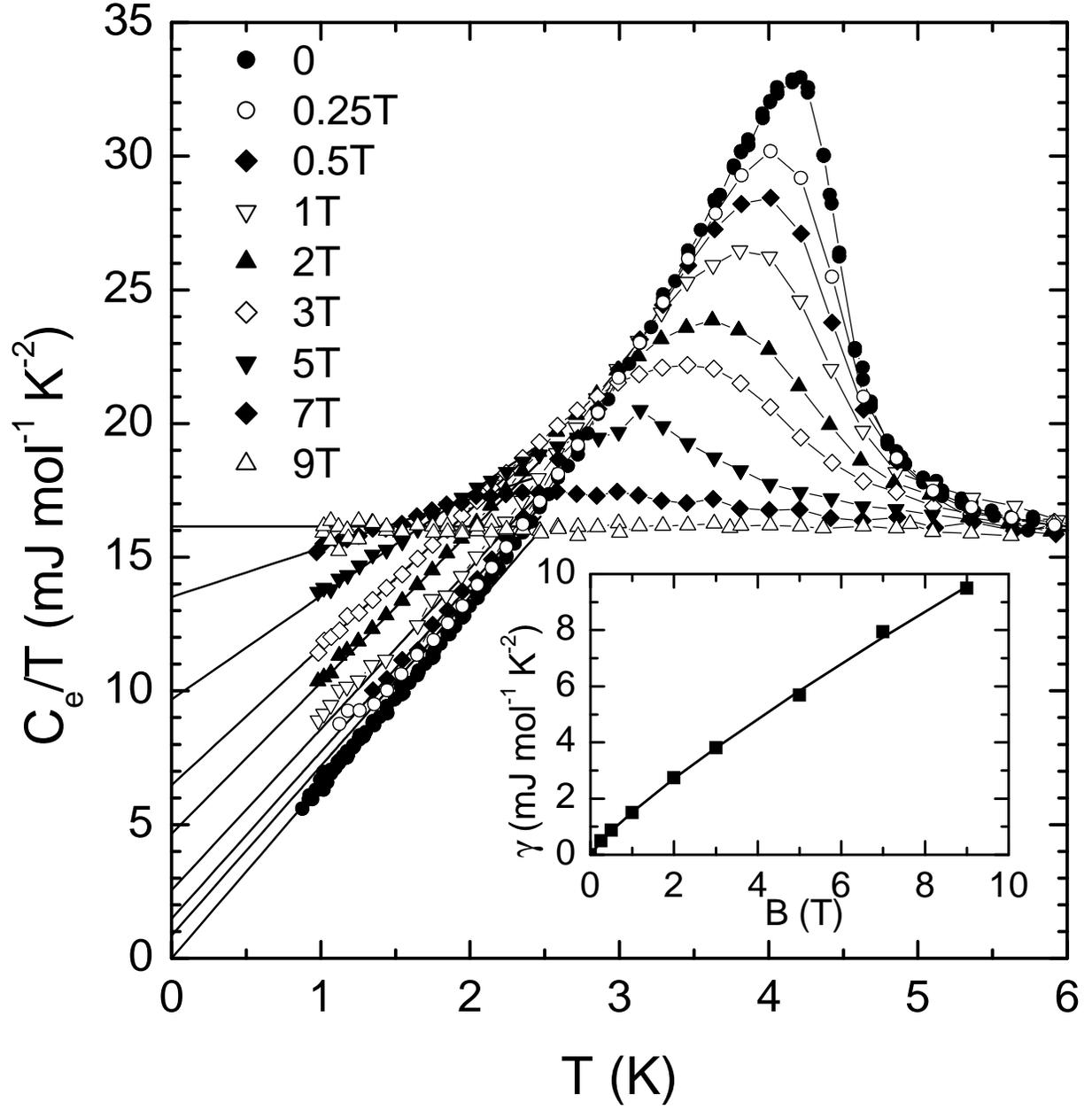}
\caption{Vortex-state heat capacity of
Na$_{0.3}$CoO$_2$$\cdot$1.3H$_2$O.  The inset shows the magnetic
field dependence of the temperature-proportional
term.}\end{center}\end{figure}
\begin{figure}\begin{center}\includegraphics[angle=-90, width=\textwidth]{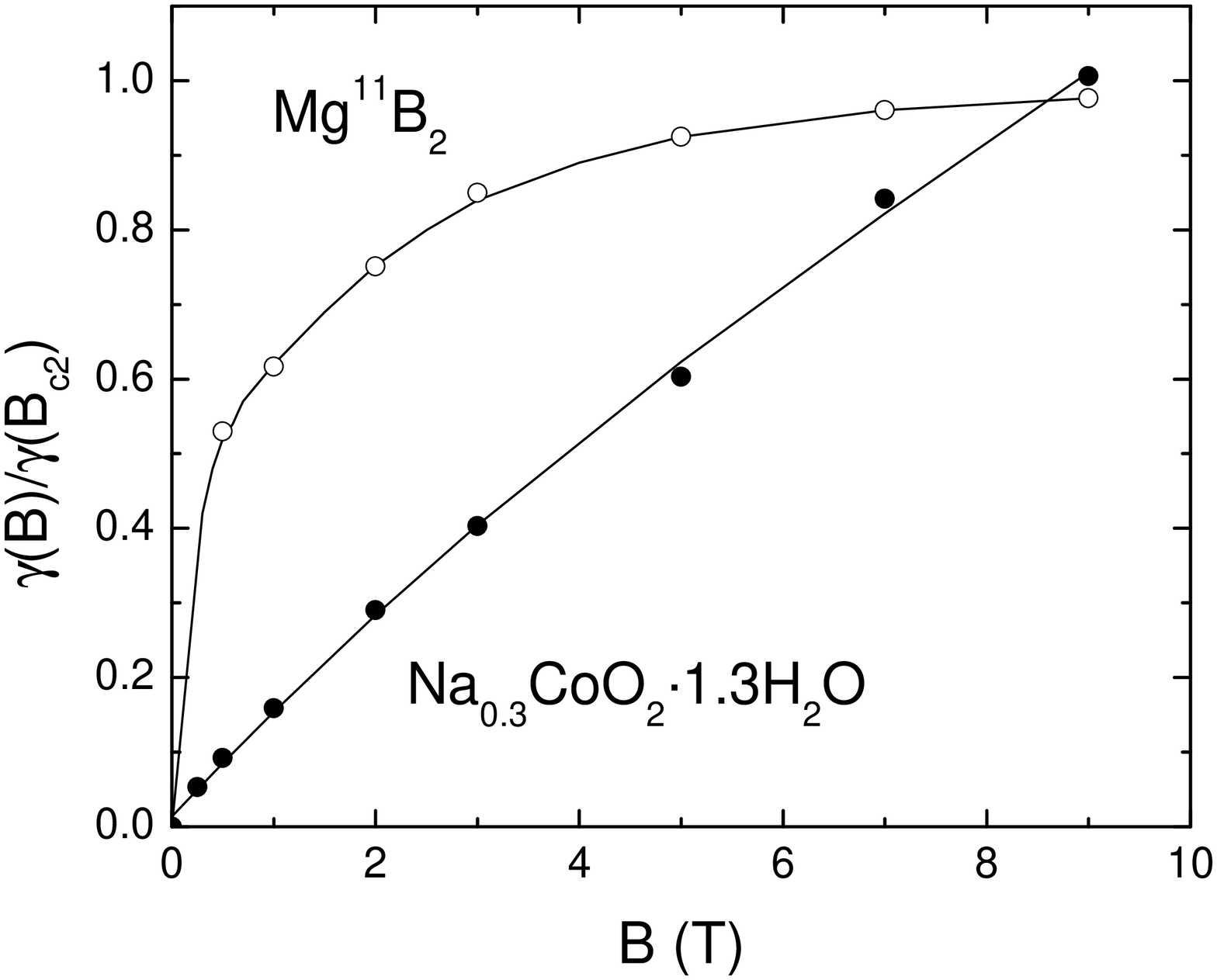}
\caption{Comparison of $\gamma$ in the vortex state for MgB$_2$
and Na$_{0.3}$CoO$_2$$\cdot$1.3H$_2$O.}\end{center}\end{figure}


\begin{thebibliography}{99}
\bibitem{nag01}J. Nagamatsu et al., Nature $\textbf{410}$, 63 (2001).
\bibitem{tak03}K. Takada et al., Nature $\textbf{422}$, 53 (2003).
\bibitem{bcs}J. Bardeen et al., Phys. Rev. $\textbf{108}$, 1175 (1957).
\bibitem{iso}S. L. Bud'ko et al., Phys. Rev. Lett. $\textbf{86}$, 1877 (2001);
D. G. Hinks et al., Nature $\textbf{411}$, 457 (2001).
\bibitem{an01}J. M. An and W. E. Pickett, Phys. Rev. Lett. $\textbf{86}$, 4633 (2001).
\bibitem{kor01}J. Kortus et al., Phys. Rev. Lett. $\textbf{86}$, 4656 (2001).
\bibitem{liu01}A. Y. Liu, I. I. Mazin, and J. Kortus,  Phys. Rev. Lett. $\textbf{87}$, 87005 (2001).
\bibitem{cho02a} H. J. Choi et al., Phys. Rev. B $\textbf{66}$, 20513 (2002).
\bibitem{cho02b} H. J. Choi et al., Nature $\textbf{418}$, 758 (2002).
\bibitem{gol}A. Golubov et al., J. Phys. Condens. Matt. $\textbf{14}$, 1353 (2002).
\bibitem{smw}H. Suhl et al., Phys. Rev. Lett. $\textbf{3}$, 552 (1959).
\bibitem{kres}B. T. Geilikman et al., Sov. Phys. Solid State $\textbf{9}$, 642 (1967);
V. Z. Kresin, J. Low Temp. Phys. $\textbf{11}$, 519 (1973); V. Z. Kresin and S. A. Wolf, Physica C $\textbf{169}$, 476 (1990).
\bibitem{spec}F. Giulibeo et al., Phys. Rev. Lett. $\textbf{87}$, 177008 (2001); P. Martinez-Samper et al., Physica C $\textbf{385}$, 233 (2003); P. Szabo et al., Phys. Rev. Lett. $\textbf{87}$, 137005 (2001); X. K. Chen et al., Phys. Rev. Lett. $\textbf{87}$, 157002 (2001); S. Tsuda et al., Phys. Rev. Lett. $\textbf{87}$, 177006 (2001).
\bibitem{bm}J. G. Bednorz and K. A. M\"{u}ller, Z. Phys. B $\textbf{64}$, 189 (1986).
\bibitem{sch}A. Schilling et al., Nature $\textbf{362}$, 226 (1993).
\bibitem{nqmr}K. Ishida et al., J. Phys. Soc. Jpn. $\textbf{72}$, 3041 (2003); T. Fujimoto et al., Phys. Rev. Lett. $\textbf{92}$, 47004 (2004).
\bibitem{kan}A. Kanigel et al., Phys. Rev. Lett. $\textbf{92}$, 257007 (2004).
\bibitem{gap}Y. Kobayashi et al., J. Phys. Soc. Jpn. $\textbf{72}$, 2161 (2003); T. Waki et al., cond-mat/0306036.
\bibitem{sph1}R. Jin et al, Phys. Rev. Lett. $\textbf{91}$, 217001 (2003); G. Cao et al., J. Phys. Condens. Matt. $\textbf{15}$, L519 (2003); B. Lorenz et al., Physica C $\textbf{402}$, 106 (2004); B. G. Ueland et al., Physica C $\textbf{402}$, 27 (2004); F. C. Chou et al., Phys. Rev. Lett. $\textbf{92}$, 157004 (2004).
\bibitem{sph2}H. D. Yang et al., cond-mat/0407589.
\bibitem{cmgb}F. Bouquet et al., Phys. Rev. Lett. $\textbf{87}$, 47001 (2001); H. D. Yang et al., \textbf{87}, 167003 (2001); Y. Wang et al., Physica C $\textbf{355}$, 179 (2001).
\bibitem{phyc}R. A. Fisher et al., Physica C $\textbf{385}$, 180 (2003).
\bibitem{2gap}F. Bouquet et al., Europhys. Lett. $\textbf{56}$, 856 (2001).
\bibitem{alph}H. Padamasee et al., J. Low Temp. Phys. $\textbf{12}$, 387 (1973).
\bibitem{foo}M. L. Foo et al., Solid State Commun. $\textbf{127}$, 33 (2003).
\bibitem{h2o}P. V. Hobbs, Ice Physics, Clarendon, Oxford, 1974.
\bibitem{nomu}T. Nomura and K. Yamada, J. Phys. Soc. Jpn. $\textbf{71}$, 404 (2002).
\end{thebibliography}
\end{document}